\documentstyle[preprint,aps,eqsecnum]{revtex}

\def\beq#1{\begin{equation} \label{#1}}
\def\eeq{\end{equation}}
\newcommand{\bea}{\begin{eqnarray}}
\newcommand{\eea}{\end{eqnarray}}
\def\bra#1{\left\langle #1\right\vert}
\def\ket#1{\left\vert #1\right\rangle}
\def\epsp{\epsilon^{\prime}}
\def\NPB{{ Nucl. Phys.} B}
\def\PLB{{ Phys. Lett.} B}
\def\PRL{ Phys. Rev. Lett.}
\def\PRD{{ Phys. Rev.} D}
\def\AJP{{\em Am. J. Phys.}}
\begin{document}
{
\tighten

\title {Simple quantum mechanics
explains GSI Darmstadt oscillations\\ Even with undetected neutrino; Momentum conservation requires \\
Same interference producing oscillations in initial and final states
}
\author{Harry J. Lipkin\,\thanks{Supported in part by
U.S.
Department of Energy, Office of Nuclear Physics, under contract
number
DE-AC02-06CH11357.}}
\address{ \vbox{\vskip 0.truecm}
  Department of Particle Physics
  Weizmann Institute of Science, Rehovot 76100, Israel \\
\vbox{\vskip 0.truecm}
School of Physics and Astronomy,
Raymond and Beverly Sackler Faculty of Exact Sciences,
Tel Aviv University, Tel Aviv, Israel  \\
\vbox{\vskip 0.truecm}
Physics Division, Argonne National Laboratory,
Argonne, IL 60439-4815, USA\\
~\\harry.lipkin@weizmann.ac.il
\\~\\
}

\maketitle

\begin{abstract}

GSI experiment on K-capture decay of radioactive ion
investigates  neutrino masses and mixing without detecting neutrino.
States of neutrinos emitted in beta decay include coherent
linear combinations of states with different masses, different momenta and same
energy. Weak decay described by Fermi Golden Rule conserves momentum but not unperturbed energy.
Continuous monitoring collapses wave function and broadens decay width.
Initial state before transition also contains coherent linear combination of
states with same momentum difference, well defined relative magnitude and phase but
broadened energies.
One-particle state with a definite momentum
difference also has an easily calculated energy difference.
Short time between last monitoring and decay allows broadened initial states with different unperturbed energies to decay to final states with single energy.
In time interval between creation of ion and decay a linear combination of two states with different unperturbed energies oscillates in time. Measuring oscillation period gives value for difference between squared neutrino masses of two neutrino mass eigenstates.
Value obtained from crude approximation with no free parameters for this `two-slit" 
experiment in momentum space differs by less than 10\%
from result observed by KAMLAND.  Observing only ion disappearance without detecting  neutrino avoids  signal suppression by low neutrino absorption cross section.

\end{abstract}

} 


\def\beq#1{\begin{equation} \label{#1}}
\def\eeq{\end{equation}}
\def\bra#1{\left\langle #1\right\vert}
\def\ket#1{\left\vert #1\right\rangle}
\def\epsp{\epsilon^{\prime}}
\def\NPB{{ Nucl. Phys.} B}
\def\PLB{{ Phys. Lett.} B}
\def\PRL{ Ph
ys. Rev. Lett.}
\def\PRD{{ Phys. Rev.} D}
\section{Introduction}

\subsection {The wave function of an unobserved neutrino}

A recent experiment\cite{gsi} describes an oscillation observed in the decay of
a radioactive ion before and during the emission of an unobserved neutrino.
This phenomenon offers a new and very interesting method for determining
neutrino masses and mixing angles\cite{gsikienle,gsifaber,gsihjl}.

To understand oscillations in the time of decay of the initial state, we use energy-momentum conservation
to determine the properties of the initial state. Even though the neutrino itself is unobserved the fact that we know it can oscillate tells us that the final state contains a coherent mixture of neutrino mass
eigenstates with the same energy and different momenta that is  emitted in electron capture decays  and called an electron neutrino. Since momentum is conserved in the weak transition that creates the neutrino, the initial ion state must also contain a coherent mixture of two states with the same momentum difference. This property of the initial state is completely independent of whether the neutrino is detected.
One-particle ion states with different momenta have different unperturbed energies. But the initial state is repeatedly monitored showing that it has not yet decayed. The time interval between the last evidence for the ion initial state  and its decay time is so short that states with different unperturbed energies can decay into the same final state with a single energy.
The relative phase between two states with different unperturbed energies changes with time and produces the  observed oscillations.

We now calculate this energy difference and the period of oscillations.

\subsection{Momentum conservation determines path from oscillating $\nu$ to mother ion}

The weak decay is a transition between the initial ``mother" ion wave packet to a final state containing a recoil ``daughter" ion with a definite energy and momentum and a neutrino wave packet which contains states with different neutrino mass, different momenta and the same energy. This decay is described by first order time-dependent perturbation theory\cite{LipQM}. The
transition from an initial state denoted by $\ket{i(t)}$ to a given final state denoted by $\ket{f}$ is given by
Fermi's Golden
Rule.
The transition probability per unit time at time $t$ is
\beq{fermi}
W(t) = \frac{2\pi}{\hbar}|\bra{f} T \ket {i(t)}|^2\rho(E_f)
\eeq
where $\bra{f} T \ket {i(t)}$ denotes transition matrix element determined in this case by weak interaction theory.
This treatment shows that momentum is conserved. The result that energy conservation is violated at short times is confirmed experimentally by the broadening of decay widths at short time. This broadening is important for the understanding of the oscillations.

Consider a component of the initial mother wave  packet which has a momentum $\vec P$ and energy $E_i$.
The final state has a recoil ion with momentum denoted
by $\vec P_R$ and  energy $E_R$ and a neutrino with mass $m$, energy $E_\nu$
and momentum  $\vec p_\nu$.  We assume conservation of momentum, but that energy is not
conserved because of the short time between the last monitoring and observation of the final
state. The energy of the final state is denoted by $E_f \not= E_i$. The conservation laws then require

\beq{epcons2} E_R= E_f - E_\nu;  ~ ~ \vec P_R = \vec P - \vec p_\nu ; ~ ~
(E_f - E_\nu)^2 -(\vec P - \vec p_\nu)^2=M_R^2
\eeq
\beq{epcons2b}
E_f^2 + E_\nu^2 -\vec P^2 - \vec p_\nu^2=E_f^2 - E_i^2 + M^2 + m^2 =M_R^2 + 2E_f E_\nu - 2\vec P\cdot\vec p_\nu
\eeq
\beq{delm2a}
\Delta (m^2) \approx (E_i+ E_f)( \delta E_i-\delta E_f) + 2E_f \delta E_\nu - 2 P \delta p_\nu
\approx  E_f \delta E_i + (E_f-E_i)\delta E_f -  2P (\delta P)
 \eeq
\beq{delm2g}
\frac{\Delta (m^2)}{ 2P \delta P}=\frac{\Delta (m^2)}{2E_i\delta E_i}\approx  \frac {E_f}{2E_i} + \frac{(E_f-E_i)\delta E_f}{2E_i\delta E_i} - 1\approx \frac {E_f-E_i}{2E_i} \cdot \left[{1}+\frac{\delta E_f}{\delta E_i}\right] - \frac{1}{2}\approx - \frac{1}{2}
 \eeq
Where we have noted that the small violation of energy conservation $(E_f-E_i)\ll E_i$
\subsection{The period of oscillation}

The phase difference at a time t between states produced by the neutrino
mass difference on the motion of the initial ion
in the laboratory frame is
\beq{delphipotalt}
\delta \phi \approx - \delta E_i \cdot t =\frac{\Delta (m^2)}{E_i}=\frac{\Delta (m^2)}{\gamma M}
\eeq
where $\gamma$ denotes the Lorentz factor $E/M$.
The period of oscillation $\delta t$ is obtained by setting  $\delta \phi \approx -2\pi$,
\beq{deltat}
\delta t \approx  {{2 \pi  E_i}\over{\Delta (m^2)}}={{2 \pi \gamma M}\over{\Delta (m^2)}}
\eeq

The  previously obtained\cite {gsikienle} theoretical value for $\Delta (m^2)$, denoted by $\Delta (m^2)_{Kienle}$, differs from ours (\ref{deltat}) denoted by $\Delta (m^2)_{HJL}$

\beq{delm2potx}
\Delta (m^2)_{Kienle}= {{4 \pi \gamma M}\over{\delta t}} \approx
2.75 \Delta (m^2)_{exp}; ~ ~ ~
\Delta (m^2)_{HJL}\approx \frac{\Delta (m^2)_{Kienle}}{2}
\approx 1.37 \Delta (m^2)_{exp}
\eeq
where the value of $\Delta (m^2)_{exp}$ is the value obtained from neutrino
oscillation experiments\cite{gsikienle}

That our theoretical value for $\Delta (m^2)$ obtained with minimum assumptions and no
free parameters is so close to the experimental value obtained from completely
different experiments suggests that better values obtained from better
calculations can be very useful in determining the masses and mixing angles
for neutrinos.
\section{Details of The K-capture experiment}

A radioactive nucleus in an ion decays by capturing an electron  from the
K-shell or other atomic shell and emits a monoenergetic  neutrino.
The emitted electron-neutrino $\nu_e$ is a linear combination
of several neutrino mass eigenstates.  If the initial state has a definite
momentum and energy and if energy and momentum are conserved,  the
energy and momentum of the neutrino are determined and therefore its mass. This would then
be a missing mass" experiment in which the mass of the neutrino is determined
without the observation of the neutrino. Interference between amplitudes from
different neutrino mass states cannot be observed in such a missing-mass
experiment. The experimental observation that interference actually occurs shows that
this cannot be a missing mass experiment. Energy is not
conserved because of the short time between the last time when the ion was observed to have
not yet decayed and the decay time.

It may seem rather  peculiar that neutrino oscillations can be observed
in the state of a radioactive ion before its decay into an unobserved neutrino.
One wonders about causality and how the initial ion can know how it will decay.
But much discussion and thought revealed that the essential quantum mechanics
is a ``two-slit" or ``which-path" experiment\cite{leofest} in momentum-energy space.
Causality is preserved because no information about the final state is
available to the initial ion.
\subsection {Actual measurement in observation of decay is not generally understood}
\begin{itemize}
\item {The ion is monitored at regular intervals during passage around the storage ring.}
\item Each monitoring collapses the wave function (or destroys entanglement phase).

\item Time in the laboratory frame is measured at each wave function collapse.

\item The the decay of the initial state is observed by the disappearance of the ion between successive
monitorings.
        \end{itemize}
         Repeated monitoring by interactions with laboratory environment at regular time
intervals and same space point in laboratory collapses wave function
and destroys entanglement\cite{Zoltan}
First-order time dependent perturbation theory gives probability for initial
state decay during small interval between two monitoring events. Final amplitudes
completely separated at long times have broadened energy spectra overlapping at short times. Their
interference produces oscillations between Dicke superradiant\cite{Super} and subradiant states having different
transition probabilities.

Experiment measures momentum difference
between two contributing coherent initial states and obtains information about $\nu$ masses without
detecting $\nu$. Simple model relates observed oscillation to squared $\nu$ mass difference and gives value
differing by less than 50\% from values calculated from KAMLAND experiment.
Monitoring simply expressed in laboratory frame not easily transformed to other frames and missed in
Lorentz-covariant descriptions based  on relativistic quantum  field theory.

The initial ion wave function is a wave packet containing a combination of
energies and momenta. The weak decay transition then produces a recoiling ion
and a final neutrino in a coherent mixture of its mass eigenstates.
It is not a missing mass experiment because the energies of the components of the
initial state wave function were not measured and nothing about the final neutrino
state was measured.
\subsection {Oscillations are produced on an initial state even without
neutrino detection}
\begin{enumerate}
\item Oscillations in time can occur only if there is interference between two components within
the initial state wave function with different energies.
\begin{itemize}
 \item The initial state has a wave function with a definite mass
 \item Interferent between components of the initial wave function with different energies
must have different momenta.
\item If energy and momentum are conserved in the transition the final state must also have
components with both different energies and different momenta.
\end{itemize}
\item Which components of the unmeasured final state are coherent?
\begin{itemize}
\item In ordinary neutrino oscillations the detector chooses coherent components
with the same energy and different momenta
\item Here there is no detector. Any coherent final state must be mixturew of
both energy and momentum
\end{itemize}
\item Coherence occurs between components of the final $\nu_e$ state with different masses,
momenta and energies but the same velocity.
\begin{itemize}
\item Components of a $\nu_e$ state with different masses,
momenta and energies but the same velocity remain a  $\nu_e$ state forever
\item Coherence arises when components of an initial state have the same momentum
and energy differences as the components a  $\nu_e$ state having the same velocity
\end{itemize}
\end{enumerate}

Since the same final $\nu_e$ state can be produced by any of the momentum components in
the initial wave function, the path in energy-momentum space between the
initial and final states is not known and the corresponding amplitudes can be
coherent and interfere.

The relative phases in the initial wave function are determined by its
localization in space at the point of entry into the apparatus. These relative
phases change with time in accordance with the relative energy differences in
the packet. They are independent of the final state, which is created only at
the decay point. Thus there is no violation of causality.
$\nu_e$ from several mass eigenstates depends upon the relative
phases of the contributions from components in the inital wave function having
different energies and momenta. These relative phases increase linearly with
time and produce oscillations.

Observing the period of these oscillations
gives information about the neutrino mass differences and the mixing angles of
the neutrino mass matrix. Reliable detailed values for the relation between the
observed oscillation period and neutrino mass differences are not obtained in
the crude models so far considered.   At this point the fact that the value
obtained (\ref{deltat}) is so close to
values obtained from neutrino
oscillation experiments is encouraging.

\subsection{Dicke superradiance and subradiance in the experiment}

The initial radioactive ``Mother" ion is in a one-particle state with a
definite mass moving in a storage ring. There is no entanglement\cite{Zoltan}
since no other particles are present.
The final state denoted by $\ket{f(E_\nu)}$ has a ``daughter" ion and an electron neutrino $\nu_e$
which is a linear  combination of  two
neutrino mass eigenstates denoted by $\nu_1$ and $\nu_2$  with masses $m_1$ and
$m_2$. To be coherent and produce oscillations the two components of
the final wave function must have the same energy $E_\nu$ for the neutrino
and
the same momentum $\vec P_R$ and  energy $E_R$ for the ``daughter" ion.
\beq{final2com}
\ket{f(E_\nu)}\equiv \ket{\vec P_R;\nu_e(E_\nu)}  =
\ket{\vec P_R;\nu_1(E_\nu)}\bra{\nu_1}\nu_e\rangle + \ket{\vec P_R;\nu_2(E_\nu)}\bra{\nu_2}\nu_e\rangle
\eeq
where  $\bra{\nu_1}\nu_e\rangle$ and $\bra{\nu_2}\nu_e\rangle$ are elements
of the neutrino mass mixing matrix, commonly expressed in terms of a
mixing angle denoted by $\theta$.
\beq{final3com}
\cos \theta \equiv \bra{\nu_1}\nu_e\rangle; ~ ~ ~
 \sin \theta \equiv \bra{\nu_2}\nu_e\rangle; ~ ~ ~\ket{f(E_\nu)}
= \cos \theta \ket{\vec P_R;\nu_1(E_\nu)}+ \sin \theta \ket{\vec P_R;\nu_2(E_\nu)}
\eeq

We use a simplified two-component initial state for the ``mother" ion
having two components
$\ket{\vec P,E}$ and $\ket{(\vec P +\delta \vec P),(E +\delta E)}$.
 Since the states $\nu_1(E_\nu)$ and $\nu_2(E_\nu)$ have the same energies and different masses, they
have different momenta.
After a very short time two components with different initial
state energies can decay into a final state which has two components with the
same energy and a
neutrino state having two components with the same momentum difference
$\delta \vec P$ present in the initial state.

The momentum conserving transition matrix elements between the two initial
momentum components to final states with the same energy and momentum difference
$\delta \vec P$ are
\beq{transcom}
\bra{f(E_\nu)} T \ket {\vec P)} = \cos \theta \bra {\vec P_R;\nu_1(E_\nu)}T \ket {\vec P)}
;~ ~ ~
\bra{f(E_\nu)} T \ket {\vec P + \delta \vec P)} =\sin \theta \bra {\vec P_R;\nu_2(E_\nu)}T
\ket {\vec P + \delta \vec P)}
\eeq

The Dicke superradiance\cite{Super} analog here is seen by defining
superradiant and subradiant linear combinations of these states
\beq{super}
\ket{Sup(E_\nu)}\equiv
\cos \theta \ket {P)} + \sin \theta \ket {P + \delta P)}; ~ ~ ~
\ket{Sub(E_\nu)}\equiv \cos \theta \ket {P + \delta P)}- \sin \theta \ket {P)}
\eeq
The transition matrix elements for these two states are then
\beq{trans}
\frac{\bra {f(E_\nu)} T \ket {Sup(E_\nu)}}{\bra{f} T \ket {P }} =[\cos \theta +
\sin \theta ]
; ~ ~ ~
\frac{\bra {f(E_\nu)} T \ket {Sub(E_\nu)}}{\bra{f} T \ket {P }} =  [\cos \theta -
\sin \theta ]
\eeq
where we have neglected the dependence of the transition operator $T$ on the
small change in the momentum $P$.
The squares of the transition matrix elements are

\beq{transsupsubsq}
\frac{|\bra {f(E_\nu)} T \ket {Sup(E_\nu)}|^2}{|\bra{f} T \ket {P }|^2} =
[1 + \sin 2 \theta ]
; ~ ~ ~
\frac{|\bra {f(E_\nu)} T \ket {Sub(E_\nu)}|^2}{ |\bra{f} T \ket {P }|^2 }=
[1 - \sin 2 \theta ]
\eeq

For maximum neutrino mass mixing, $\sin 2 \theta =1$ and
\beq{transsupsubmax}
|\bra {f(E_\nu)} T \ket {Sup(E_\nu)}|^2 =
2 |\bra{f} T \ket {P }|^2
; ~ ~ ~
|\bra {f(E_\nu)} T \ket {Sub(E_\nu)}|^2 = 0
\eeq

This is the standard Dicke superradiance in which all the transition strength
goes into the  superradiant state and there is no transition from the
subradiant state.

Thus from eq.
(\ref{super}) the initial state at time t varies periodically between the
superradiant and
subradiant states.

\subsection {How a ``watched pot experiment" can give a nonexponential decay}

The experiment observes a time-dependence in the decay probability which is not
exponential. Although this appears at first to be counterintuitive, it follows
naturally from
a crucial feature of being a ``watched pot" experiment. The initial state
of the ion is monitored during its passage around a storage ring, thereby
affirming that the ion has not yet decayed.  It is like the
``Schroedinger cat" experiment in which the door is always open so that there
is a continuous measurement of whether the cat is still alive.

The wave function describes the motion of the initial state as a free ion moving
in the fields of the apparatus for a time $t$ defined as the time interval
between its entry into the apparatus and the last time before the decay in which
it was affirmed not to have decayed. The time $t'$ between the last monitoring
and the time of decay is negligible for the purpose of the analysis of the
experiment.
\beq{twotimes}
t' \ll t
\eeq
The initial state denoted by $\ket {i}$ is a wave packet containing components
with different energies. The relative phases of these components in the initial wave function
are determined by its  localization in space at the point of entry into the
apparatus. The changes with time of these relative phases are described by a
Hamiltonian denoted by $H_o$ which describes the motion of a free initial ion
moving in the
electromagnetic fields constraining its motion in a storage ring. The wave
function describing the evolution of the initial state in time is thus
\beq{timedep}
\ket{i(t)} = e^{iH_ot} \ket {i}
\eeq
The decay transition to a given final state denoted by $\ket{f}$ is described by
a transition matrix element
\beq{trans}
\bra{f} T \ket {i(t)} = \bra{f} T e^{iH_ot} \ket {i}
\eeq
The transition probability per unit time at time $t$ is given by Fermi's Golden
Rule,
\beq{fermi}
W(t) = \frac{2\pi}{\hbar}|\bra{f} T \ket {i(t)}|^2\rho(E_f) =
\frac{2\pi}{\hbar}|\bra{f} T e^{iH_ot} \ket {i}|^2\rho(E_f)
\eeq
We can now see why the time dependence of the decay is not exponential.
The probability $P_i$ that the ion is still in its initial state and has not
yet decayed satisfies the differential equation
\beq{diffeq}
\frac {d}{dt} P_i = - W(t) P_i; ~ ~ ~ ~ ~
\frac {d}{dt} log (P_i) = - W(t).
\eeq
Solving this equation gives
\beq{nonexp}
P_i = e^{-\int W(t)dt}
\eeq
We see immediately why decays are generally exponential and this one need not
be. Usually the transition matrix element
(\ref{trans}) and the transition probability (\ref{fermi}) are independent of
time and eq. (\ref{nonexp}) gives an exponential decay. Here the transition
probabilty depends upon the propagation of the initial state during the time
$t$ between the entry of the ion into the apparatus and the time of the decay.

For a simple gedanken example consider the decay of a spin-1/2 radioactive
particle with a spin-dependent interaction which allows it to decay with a
lifetime of five days but only when the spin is polarized in the $+x$
direction. The decay is forbidden from the $-x$ state. Consider a beam of such
particles moving in the z-direction, polarized at time $t=0$ in the $+x$
direction with a weak magnetic field in the z-direction causing the spin to
precess with a period of seven seconds around the $z$-axis. The decay rate
will not be exponential but will be modulated by a periodic function with a
period of seven seconds.

Since the time dependence depends only on the propagation of the initial state,
it is independent of the final state, which is created only at the decay point.
Thus there is no violation of causality. No information about the final state
exists before the decay.  Although time-dependent perturbation theory might
suggest that a decay amplitude can be present before the decay, the continued
observation of the initial ion before the decay rules out any influence of any
final state amplitude on the decay process.

\subsection{A tiny energy scale}
The experimental result, if correct, sets a scale  in time  of seven
seconds, which means a tiny energy scale for the difference between two waves
which beat with a period of seven seconds.
\beq{beat}
\Delta E \approx 2\pi \cdot \frac {\hbar}{7}
= 2\pi \cdot \frac {6.6\cdot 10^{-16}}{7} \approx 0.6\cdot 10^{-15}{\rm eV}
\eeq
This tiny energy scale cannot come out
of thin air. It must be predictable from standard quantum mechanics  using a
scale from another input. The only other input available according to eq.
(\ref{nonexp}) is in the propagation of the initial state through the storage
ring during the time interval between the entry into the apparatus and the
decay. One tiny scale available in the parameters
that describe this experiment is the mass-squared diffgference between two
neutrino mass eigenstates. This gives a tiny mass scale when this mass-squared
is divided by the energy of the ion.
\beq{mscale}
\frac{\Delta (m^2)}{E} \approx \frac {0.8\cdot 10{-4}}{3\cdot 10^{11}} \approx
2.7\cdot 10^{-15}{\rm eV}
\eeq
where the value of $\Delta (m^2)$ is obtained from neutrino
oscillation experiments\cite{gsikienle}.

That these two tiny energy scales obtained from completely different inputs are
within an order of magnitude of one another suggests that they must be related
by a serious quantum-mechanical calculation.  The simplest
model relating these two tiny mass scales gives a result that
differs only by 10\%. The fact that the observed seven second
period creates a tiny mass scale and that no other energy in this experiment
comes even orders of magnitude close to this scale suggests that this is not an
accident. These two scales appear in the analysis  of the same experiment where
there must be a theoretical prediction for the seven second scale if we believe
quantum mechanics.

There are many other possible mechanisms for producing oscillations. The
experimenters\cite{gsi} claim that they have investigated all of them. We also
note that all these other mechanisms involve energy scales very different from
the scale producing a seven second period.

\section{The two principal difficulties of neutrino experiments}

\begin{enumerate}
\item Ordinary neutrino oscillation experiments are difficult because
\begin{itemize}
 \item The neutrino absorption cross section is tiny. The number of neutrino
 events actually used in ordinary experiments is many orders of magnitude smaller than the
 number events creating the neutrinos.
 \item The oscillation wave lengths are so large that it is difficult to
actually follow even one oscillation period in any experiment.
\end{itemize}
\item
This experiment opens up a new line for
dealing with these difficulties
\begin{itemize}
\item The oscillation is measured without detecting the neutrino.
Detection of every neutrino creation event
avoids the losses from the low neutrino absorption cross section.
   \item The long wave length problem is solved by having the
radioactive source move a long distance circulating around in a storage
ring.  The data if
correct show many oscillations in the same experiment.

\end{itemize}
\end{enumerate}

 This paper considers the  basic quantum mechanics of the first difficulty and
shows in a crude approximation that it is possible in principle to observe and measure
neutrino oscillations by looking only at the radioactive source.

The theoretical analysis in this paper was motivated by discussions with  Paul
Kienle at a very early stage of the experiment in trying to understand whether
the effect was real or just an experimental error.

\section{Conclusions}

A new oscillation  phenomenon providing information about neutrino mixing is
obtained by following the initial radioactive ion  before and during the decay.
The difficulties introduced in conventional neutrino experiments by the tiny
neutrino absorption cross sections and the very long oscillaton wave lengths
are avoided here. Measuring the decay time enables every neutrino event to be
observed and counted without the necessity of observing the neutrino via the
tiny absorption cross section. The confinement of the initial ion in a storage
ring enables long wave lengths to be measured within the laboratory.

Coherence between amplitudes produced by the weak decay of a radioactive  ion
by the emission of  neutrinos with different masses has been shown to follow
from the localization of the initial radioactive ion within a space
interval much  smaller than the oscillation wave length.  This coherence is
observable in following the motion of the initial radioactive ion from its
entry into the apparatus to its decay.
\section{Acknowledgement}

It is a pleasure to thank Paul Kienle for calling my attention to this problem
at the Yukawa Institute for Theoretical Physics at Kyoto  University, where
this work was initiated during the YKIS2006 on ``New  Frontiers on QCD".
Discussions on possible experiments with Fritz Bosch, Walter Henning, Yuri
Litvinov and Andrei Ivanov are also gratefully acknowledged along with a
critical review of the present manuscript. The author also acknowledges further
discussions on neutrino oscillations as ``which path" experiments with  Eyal
Buks, Avraham Gal, Terry Goldman, Maury Goodman,  Yuval Grossman, Moty Heiblum, Yoseph Imry,
Boris Kayser, Lev Okun, Gilad Perez, Murray Peshkin, David Sprinzak, Ady Stern,
Leo  Stodolsky and Lincoln Wolfenstein,

%
\catcode`\@=11 
\def\references{
\ifpreprintsty \vskip 10ex
%
\hbox to\hsize{\hss \large \refname \hss }\else
\vskip 24pt \hrule width\hsize \relax \vskip 1.6cm \fi \list
{\@biblabel {\arabic {enumiv}}}
{\labelwidth \WidestRefLabelThusFar \labelsep 4pt \leftmargin \labelwidth
\advance \leftmargin \labelsep \ifdim \baselinestretch pt>1 pt
\parsep 4pt\relax \else \parsep 0pt\relax \fi \itemsep \parsep \usecounter
{enumiv}\let \p@enumiv \@empty \def \theenumiv {\arabic {enumiv}}}
\let \newblock \relax \sloppy
 \clubpenalty 4000\widowpenalty 4000 \sfcode `\.=1000\relax \ifpreprintsty
\else \small \fi}
\catcode`\@=12 
{\tighten

}
\end{document}